\documentclass[11pt]{iopart}

\newcommand\be{\begin{equation}}
\newcommand\ee{\end{equation}}
\newcommand\bea{\begin{eqnarray}}
\newcommand\eea{\end{eqnarray}}

\usepackage{setstack}
\usepackage{iopams}

\begin{document}

\title{Thermodynamics of Anomaly-Driven Cosmology}

\author{James E. Lidsey\footnote{Email: J.E.Lidsey@qmul.ac.uk}}

\address{Astronomy Unit, School of Mathematical Sciences,\\
Queen Mary University of London, London E1 4NS, United Kingdom}

\begin{abstract}
The Friedmann equations of general relativity can be derived 
from the first law of thermodynamics when the entropy of the 
apparent horizon of a spatially isotropic universe is given by 
the Bekenstein-Hawking entropy. We point out that 
if the entropy of the apparent horizon receives a logarithmic correction, 
the first law 
of thermodynamics leads to a modified Friedmann equation which corresponds 
precisely to the time-time component of the semi-classical 
Einstein field equations sourced by the trace anomaly of 
${\cal{N}}=4$ $U(N)$ super-Yang-Mills theory. 
This correspondence allows for a thermodynamic description of the dynamics
of the Randall-Sundrum braneworld scenario. 
\end{abstract}

\vskip 1pc

\section{Introduction}

It is well known 
that the entropy and temperature of the event 
horizon of a Schwarzschild black hole 
are given by the Bekenstein-Hawking entropy \cite{bekhawk}  
\begin{equation}
\label{bekhawk}
S_{\rm BH}=\frac{A}{4G} 
\end{equation} 
and the Hawking temperature \cite{hawk}
\begin{equation}
\label{hawktemp}
T_H =\frac{1}{2\pi \tilde{r}_A} ,
\end{equation}
respectively, where $A= 4\pi \tilde{r}_A^2$ 
is the proper area of the horizon 
and $G$ is Newton's constant. Similar 
expressions arise for the cosmological event horizon of de Sitter
space \cite{gibhawk}. 
More generally, if the entropy of a local Rindler causal horizon 
is proportional to its area, the Einstein field equations 
can be derived from the fundamental thermodynamic relation 
$\delta Q = TdS $, where the heat flow, $\delta Q$, 
is interpreted in terms of the energy flux 
across the horizon and the temperature is the Unruh 
temperature associated with a uniformly accelerating observer 
just within the horizon \cite{jac}. 
Similar ideas have been employed in a cosmological context
\cite{frokof,dan,bousso,wang,gongwang,caikim,wwy}. It can be 
shown that if the entropy of the apparent horizon of a spatially isotropic 
universe is related to its area by Eq. (\ref{bekhawk}), 
the Friedmann equations of general relativity follow  
from the first law of thermodynamics \cite{caikim}. 

These developments indicate that there exists a direct 
correspondence between Einstein gravity and   
the Bekenstein-Hawking entropy. However,    
Eq. (\ref{bekhawk}) is a semi-classical relation 
and becomes modified when quantum corrections are taken into account.
One such correction arises when one-loop effects break the 
conformal invariance of classical massless fields and induce an 
anomalous trace for the energy-momentum tensor. 
(For reviews, see, e.g., Refs. \cite{duff,birrell}). In particular, 
the entropy of a Schwarzschild black hole that is emitting 
zero-mass particles receives a logarithmic correction of the form
\cite{furs,bm1,bm2}: 
\begin{equation}
\label{quantumentropy}
S= \frac{A}{4G} + \alpha \ln \left( \frac{A}{4G} \right)  ,
\end{equation}
where the dimensionless parameter, $\alpha$, is determined by 
the conformal anomaly of the fields.  

In a cosmological context, it was recently shown that 
a logarithmic correction to the entropy of the apparent horizon 
generates a modified Friedmann equation 
when the first law of thermodynamics is 
applied \cite{cch}. On the other hand, 
the conformal anomaly is purely 
geometrical in nature and also modifies the 
dependence of the Friedmann equation 
on the energy density of the universe. 
In view of the above correspondence between 
gravity and thermodynamics, it is of interest to investigate whether 
the modifications due to the trace anomaly can be reproduced by 
a logarithmically-corrected Bekenstein-Hawking entropy. 
In the present work, we find 
that the corrections to the spatially flat Friedmann equation 
derived from the thermodynamic and field-theoretic 
perspectives are identical when the conformal field theory (CFT) 
is ${\cal{N}} =4$, $U(N)$ super-Yang-Mills theory. As in the 
black hole background, the logarithmic correction to the 
entropy is proportional 
to the conformal anomaly of the fields. In this sense, therefore, the 
conformal anomaly can be 
interpreted as a quantum correction to the entropy of the apparent 
horizon. 

The structure of the paper is as follows. We begin in Section 2 
with the thermodynamic derivation of the Friedmann equation
and establish a direct correspondence with 
anomaly-driven cosmology in Section 3. We employ this correspondence in 
Section 4 to present a thermodynamic description of the Randall-Sundrum   
braneworld scenario. We conclude with a summary in Section 5.
 
\section{Thermodynamic Friedmann Equation}

The spatially flat and isotropic Friedmann-Robertson-Walker 
(FRW) line-element is given by  
$ds^2 = h_{ab}dx^a dx^b + \tilde{r}^2 d\Omega_2^2$, where 
$h_{ab} = {\rm diag} (-1, a^2)$, $\tilde{r} \equiv ra (t)$ 
and $a$ denotes the 
scale factor. On a constant time 
hypersurface, the apparent horizon of an observer at $r=0$ 
is the sphere where orthogonal ingoing, future-directed 
light-rays have zero expansion. This sphere is defined by the 
condition $h^{ab} \partial_a \tilde{r}
\partial_b \tilde{r} =0$ and has a radius 
$\tilde{r}_A =r_A a =1/H$ and area $A= 4\pi/H^2$, 
where $H \equiv \dot{a}/{a}$. 

We consider a universe sourced by a 
perfect fluid with diagonal energy-momentum tensor, 
$T_{\mu\nu} = {\rm diag} (\rho, a^2p, a^2p, a^2p)$, 
where $\rho$ and $p$ denote the energy 
density and pressure of the fluid, respectively. 
We assume that during an infinitesimal time interval, $dt$, the area of the 
apparent horizon is effectively constant.  
The amount of energy, $dE$, that crosses 
the horizon during this interval can then be
calculated by integrating the flux of the energy-momentum 
tensor through the horizon contracted with the (approximate) generator
of the horizon, $k^a = (1, -Hr)$. It 
follows that
\begin{equation}
\label{energyflux}
dE=  -A T_{ab}k^ak^b dt = - \frac{4\pi}{H^2} (\rho +p )dt 
= \frac{4\pi}{3}  \frac{d\rho}{H^3}  ,
\end{equation}
where the condition for conservation of energy-momentum, 
\begin{equation}
\label{conservation}
\dot{\rho} + 3H(\rho +p) =0
\end{equation}
has been employed in the last equality.  

It was recently shown that the apparent horizon of the FRW universe 
can be associated with Hawking radiation 
with a temperature given by Eq. (\ref{hawktemp}) \cite{cch2}. 
The first law of thermodynamics on the horizon, $dE= -TdS$, 
may therefore be expressed in the form 
\begin{equation}
\label{firstlaw}
dS = -\frac{8\pi^2}{3} \frac{d \rho}{H^4}  .
\end{equation} 
In the general case where the entropy of the 
apparent horizon is an arbitrary (differentiable) function of its area, 
$S=S(\pi/GH^2)$, Eq. (\ref{firstlaw}) can be 
formally integrated to yield 
\begin{equation}
\label{generalfriedmann}
\rho = -\frac{3}{8\pi^2} \int^H d{H'}^2 \left( {H'}^4 
\frac{dS}{d{H'}^2} \right)
\end{equation}
and 
Eq. (\ref{generalfriedmann}) may be interpreted as a generalized 
Friedmann equation, where the constant of integration determines 
the value of the cosmological constant\footnote{In this 
thermodynamic context, the cosmological constant is determined by the 
initial conditions \cite{dan}. We set it to zero in what follows.}. 
When the entropy is given by the Bekenstein-Hawking entropy, 
the standard Friedmann equation,  $\rho  = 3H^2/(8\pi G)$, is recovered 
\cite{caikim}. When the entropy receives a logarithmic correction of the 
form (\ref{quantumentropy}), however, integration of 
Eq. (\ref{generalfriedmann}) results in a modified Friedmann equation 
given by \cite{cch}
\begin{equation}
\label{quantumfriedmann}
H^2 + \frac{\alpha G}{2\pi} H^4 = \frac{8\pi G}{3} \rho  .
\end{equation}
Eq. (\ref{quantumfriedmann}) can be solved quadratically to yield 
\begin{equation}
\label{quadratic}
H^2 = \frac{\pi}{\alpha G} \left[ -1 + \epsilon \left( 
1+ \frac{16\alpha G^2}{3} \rho \right)^{1/2} \right]  , 
\end{equation}
where $\epsilon = \pm 1$. 

\section{Trace-Anomaly Friedmann Equation}

We now consider the effect of the conformal anomaly on 
the semi-classical gravitational field equations derived directly 
from quantum field theory. These equations
can be expressed in the form 
\begin{equation}
\label{efe}
R_{\mu\nu} -\frac{1}{2} R g_{\mu\nu}
= 8\pi G \langle T_{\mu\nu} \rangle  ,
\end{equation}
where $\langle T_{\mu\nu} \rangle$ represents the expectation value 
for the energy-momentum tensor
of a large number of conformally invariant fields. 
For a general CFT, the total one-loop contribution to the trace of this 
tensor is given by the conformal anomaly \cite{duff}: 
\begin{equation}
\label{trace}
g^{\mu\nu} \langle T_{\mu\nu} \rangle = 
c F -b G + d \nabla^2 R
\end{equation}
where $F = C_{\mu\nu\lambda\kappa}C^{\mu\nu\lambda\kappa}$ 
is the square of the Weyl tensor and 
$G = R^2 -4R^{\mu\nu}R_{\mu\nu} + 
R^{\mu\nu\lambda\kappa}R_{\mu\nu\lambda\kappa}$ is the topological Gauss-Bonnet 
invariant. The constant coefficients are determined 
by the field content of the CFT \cite{birrell}: 
\begin{eqnarray}
\label{defb}
b = \frac{1}{360 (4\pi)^2} \left( n_0 + 11n_{1/2} + 62n_1 \right)
\\
\label{defc}
c = \frac{1}{120 (4\pi)^2} \left( n_0 + 6n_{1/2} + 12n_1 \right)
\\
\label{defd}
d  = \frac{1}{180 (4\pi)^2} \left( n_0 + 6n_{1/2} -18 n_1 \right)  ,
\end{eqnarray}
where $n_s$ denotes the number of fields with spin $s$ and 
the spinor fields are Dirac fermions. 

The coefficients $b$ and $c$ are independent of the renormalization
scheme, whereas $d$ is scheme-dependent. Eq. (\ref{defd}) is the result 
derived from point-splitting or zeta-function regularization \cite{birrell}. 
The same result is predicted by the AdS/CFT correspondence 
where the CFT is ${\cal{N}} =4$, $U(N)$ super-Yang-Mills 
theory \cite{hennsken}. We consider this CFT in what follows. 
Its field content  
is given by $n_0 = 6N^2$, $n_{1/2} =2N^2$ and $n_1 = N^2$ in the 
large $N$ limit, which implies that 
\begin{equation}
\label{fieldcontent}
b=c =\frac{N^2}{(8\pi)^2} , \qquad d=0  .
\end{equation} 
Consequently, the trace of the 
one-loop energy-momentum tensor for this CFT reduces to  
\begin{equation}
\label{traceanomaly}
\langle {T^{\mu}}_{\mu} \rangle = - 24 b \frac{\ddot{a}}{a}
\frac{\dot{a}^2}{a^2}
\end{equation}
on the conformally flat and spatially flat FRW background. 

In deriving the modified Friedmann equation, 
it proves convenient to interpret the components of the quantum 
energy-momentum tensor in terms of an effective 
energy density and pressure, such that $\langle T_{00} 
\rangle  \equiv \sigma$ and 
$\langle T_{ij} \rangle  \equiv \sigma_P a^2 \delta_{ij}$
\cite{hartle,hhr2,kir}. The trace formula 
(\ref{traceanomaly}) can then be written as 
\begin{equation}
\label{sigmatrace}
\sigma -3\sigma_P =  24 b \frac{\ddot{a}}{a}
\frac{\dot{a}^2}{a^2}  ,
\end{equation}
whereas the Bianchi identity reduces to the condition 
\begin{equation}
\label{sigmaeom}
\dot{\sigma} + 3H (\sigma +\sigma_P ) =0 .
\end{equation}

Combining Eqs. (\ref{sigmatrace}) and (\ref{sigmaeom}) 
to eliminate the effective pressure implies that 
the energy density generated by 
the trace anomaly satisfies the differential equation  \cite{kir}
\begin{equation}
\label{combine}
\dot{\sigma} +4H \sigma - 24b \frac{\ddot{a}}{a} 
\frac{\dot{a}^3}{a^3} =0
\end{equation} 
and integrating Eq. (\ref{combine}) then yields the solution
\begin{equation}
\label{tracesolution}
\sigma = \frac{\sigma_0}{a^4} + 6 b H^4  ,
\end{equation}
where $\sigma_0$ is an arbitrary integration constant. 
Consequently, the $(00)$-component of the 
gravitational field equations (\ref{efe}) may now be expressed in the form
\begin{equation}
\label{tracefriedmann}
H^2 - 16\pi G b H^4 = \frac{8\pi G}{3}  \frac{\sigma_0}{a^4}  .
\end{equation}
 
We conclude after comparison with Eq. (\ref{quantumfriedmann}), 
therefore, that the Friedmann equation for  
anomaly-driven cosmology can be derived from the first law
of thermodynamics when the matter source is a conformally invariant 
classical fluid  
and the entropy of the apparent horizon 
is given by the logarithmically-corrected Bekenstein-Hawking relation 
(\ref{quantumentropy}). The correspondence is exact 
if the parameter $\alpha$ is proportional to the conformal anomaly: 
\begin{equation}
\label{propto}
\alpha = - 32 \pi^2 b = -\frac{N^2}{2}  .
\end{equation} 

In the following Section, we discuss this thermodynamic interpretation 
of trace-anomaly cosmology within the 
context of the Randall-Sundrum braneworld scenario and the AdS/CFT 
correspondence. 

\section{Randall-Sundrum Braneworld and the AdS/CFT Correspondence}

Two of the most significant advances over the past decade 
towards a consistent theory of quantum gravity have been the AdS/CFT
correspondence \cite{mald,wit,gubpoly} 
and the Randall-Sundrum (RS) braneworld scenario \cite{rs}. In 
the RS scenario, our observable four-dimensional universe is 
interpreted as a co-dimension one-brane propagating in five-dimensional 
anti-de Sitter (AdS) space with a $Z_2$ reflection symmetry. 
From the point of view of a four-dimensional
observer confined to the brane, the cosmic expansion arises due to the
propagation of the brane through the bulk space. The 
effective four-dimensional Friedmann equation can be derived 
by applying the thin wall formalism of General 
Relativity to five dimensions and it can be shown that 
for a spatially flat world-volume
\begin{equation}
\label{bulkfriedmann}
H^2 = \frac{8\pi G}{3} \rho  \left( 1 + \frac{\rho}{2\lambda} \right) ,
\end{equation}
where $\rho$ represents the density of matter on the brane,   
$\lambda =3G/(4\pi G_5^2)$ denotes the brane tension and $G_5$ 
is the five-dimensional Newton constant 
\cite{bine,cline,maeda}. (The tension is fine-tuned to this
value to ensure that the brane world-volume reduces to 
Minkowski spacetime in the vacuum limit).  
Energy-momentum conservation of the four-dimensional matter fields,
Eq. (\ref{conservation}), follows as a direct consequence of the 
vanishing of the $(\mu 5)$-component of the 
five-dimensional Einstein tensor.

The AdS/CFT correspondence provides a 
realization of the holographic principle \cite{hooft,suss} 
by relating semi-classical, $d$-dimensional gravity in 
AdS space to a quantum CFT located on the boundary of the spacetime
\cite{mald,wit}. In particular, this correspondence implies that 
the RS model can be interpreted 
holographically as four-dimensional 
Einstein gravity coupled to a CFT with an ultra-violet cutoff. 
The CFT is two copies of the ${\cal{N}}=4$ $U(N)$ super-Yang-Mills 
theory, where $ N^2 = \pi \ell^3 /(2G_5)$ and 
$\ell = G_5/G$ is the AdS radius \cite{gub,hhr}. 
It follows from Eq. (\ref{propto}), therefore, that
the Friedmann equation for this model 
is given by Eq. (\ref{generalfriedmann}), where 
$\alpha = -64\pi^2 b= -N^2$. Consequently,  
the dynamics of the holographic 
dual of the RS scenario can be described in terms of  
the first law of thermodynamics, such that
\begin{eqnarray}
\label{holographic}
dE =-TdS 
, \qquad 
T= \frac{H}{2\pi} 
\\
\label{holographicentropy}
S= \frac{\pi}{GH^2} + \alpha \ln \left( \frac{\pi}{GH^2} \right) , 
\qquad \alpha = -N^2 =  - \frac{\pi \ell^3}{2G_5} .
\end{eqnarray}

Since the four- and five-dimensional interpretations of the RS scenario 
are dual to one other, it should be possible 
to provide a thermodynamic derivation of the 
Friedmann equation (\ref{bulkfriedmann}) that followed directly from the  
five-dimensional gravitational field equations.
To address this question, we now suppose that the first law of 
thermodynamics, Eqs. (\ref{holographic})-(\ref{holographicentropy}), 
can be applied to the apparent horizon of the world-volume of the brane. 
We then proceed to express the first law in the form 
\begin{equation}
\label{reint}
TdS_{\rm BH} = -dE - TdS_{\rm cor}  ,
\end{equation}
where $S_{\rm cor} \equiv \alpha \ln ( \pi / GH^2)$ denotes the
correction to the Bekenstein-Hawking entropy, $S_{\rm BH}$.  
It follows from Eq. (\ref{firstlaw}) that Eq. (\ref{reint}) can be 
written as
\begin{equation}
\label{reexpress}
H^4 dS_{\rm BH} = -\frac{8\pi^2}{3} d\rho - 
H^4 dS_{\rm cor}   .
\end{equation}

In the absence of the logarithmic 
correction, integration of Eq. (\ref{reexpress})  
would lead to the standard Friedmann equation, 
$H^2=8\pi G\rho/3$. If we regard this as a `zero-order' solution 
to Eq. (\ref{reexpress}) and substitute it back into the  
second term on the right-hand side,
we may interpret this term as an explicit 
function of the energy density. It then follows that 
\begin{equation}
\label{iterative}
H^4 dS_{\rm BH} \simeq -   \frac{8\pi^2}{3} d\rho
- \frac{16\pi^3 \ell  G_5}{9}    d \rho^2
\end{equation}
when $\alpha$ is determined by the bulk quantities 
$\ell$ and $G_5$ given in the AdS/CFT relation (\ref{holographicentropy}). 
It is then straightforward to show that 
integration of Eq. (\ref{iterative}) leads 
directly to the RS Friedmann equation (\ref{bulkfriedmann}).
In this sense, the RS scenario may be viewed as the 
low-energy/high-entropy limit of  
the anomaly-driven cosmology. This is consistent with 
Taylor expanding Eq. (\ref{quadratic}) to quadratic order in the 
energy density \cite{kir,barv}.

\section{Conclusion}

Cosmology driven by the trace anomaly of conformally invariant matter 
fields represents an important framework for 
investigating the evolution of the very early universe. 
The correspondence highlighted in the present work 
allows us to interpret the effect of the  
anomaly from a thermodynamic perspective as 
a logarithmic correction to the Bekenstein-Hawking 
entropy associated with the apparent horizon of the spatially flat 
FRW universe. The Friedmann equation derived by applying the first law of
thermodynamics to the apparent horizon with such a corrected entropy 
has the same form as that derived directly from field-theoretic techniques. 
This is interesting given that a similar logarithmic
correction to the Bekenstein-Hawking entropy 
is generated by one-loop effects near the event
horizon of a Schwarzschild black hole \cite{furs}.
In both the cosmological and black hole backgrounds, 
the correction is determined 
by the number of zero-mass fields: 
for the FRW universe, $\alpha \propto -b$, whereas
$\alpha \propto (c-b )$ for the Schwarzschild black 
hole \cite{furs}. 

We have focused on the trace anomaly 
of ${\cal{N}} =4 $ $U(N)$ 
super-Yang-Mills theory and our conclusions rely on the condition that
the anomaly coefficient defined in Eq. (\ref{defd}) 
vanishes. A non-zero value would lead to terms 
involving higher derivatives of the scale factor
in the expression for the effective 
energy density of the anomaly, Eq. (\ref{tracesolution}). 
Nonetheless, we expect this correspondence between 
thermodynamics and gravity to hold for other theories 
where $d \ne 0$. This follows since the value of 
this parameter can always be renormalized to zero by 
introducing a finite local counterterm (proportional to 
$R^2$) into the non-local action. 

In conclusion, therefore, the present 
work provides a further example of how thermodynamics emerges from a
combination of quantum physics and gravitation.

\section*{References}

\bibliographystyle{iopart-num-ih}

\bibliography{qmw08-4c}

\providecommand{\newblock}{}
\begin{thebibliography}{10}
\expandafter\ifx\csname url\endcsname\relax
  \def\url#1{{\tt #1}}\fi
\expandafter\ifx\csname urlprefix\endcsname\relax\def\urlprefix{URL }\fi
\providecommand{\eprint}[2][]{\url{#2}}

\bibitem{bekhawk}
Bekenstein J~D, {\em Black holes and entropy\/}, 1973 {\em Phys. Rev. D\/} {\bf
  7} 2333--2346

\bibitem{hawk}
Hawking S~W, {\em Particle creation by black holes\/}, 1975 {\em Commun. Math.
  Phys.\/} {\bf 43} 43

\bibitem{gibhawk}
Gibbons G~W and Hawking S~W, {\em Cosmological event horizons, thermodynamics,
  and particle creation\/}, 1977 {\em Phys. Rev. D\/} {\bf 15} 2738--2751

\bibitem{jac}
Jacobson T, {\em {Thermodynamics of space-time: The Einstein equation of
  state}\/}, 1995 {\em Phys. Rev. Lett.\/} {\bf 75} 1260
  [\eprint{gr-qc/9504004}]

\bibitem{frokof}
Frolov A~V and Kofman L, {\em {Inflation and de Sitter thermodynamics}\/}, 2003
  {\em JCAP\/} {\bf 0305} 009 [\eprint{hep-th/0212327}]

\bibitem{dan}
Danielsson U~H, {\em {Transplanckian energy production and slow-roll
  inflation}\/}, 2005 {\em Phys. Rev.\/} {\bf D71} 023516
  [\eprint{hep-th/0411172}]

\bibitem{bousso}
Bousso R, {\em {Cosmology and the S-matrix}\/}, 2005 {\em Phys. Rev.\/} {\bf
  D71} 064024 [\eprint{hep-th/0412197}]

\bibitem{wang}
Wang P, {\em {Horizon entropy in modified gravity}\/}, 2005 {\em Phys. Rev.\/}
  {\bf D72} 024030 [\eprint{gr-qc/0507034}]

\bibitem{gongwang}
Gong Y and Wang A, {\em {The Friedmann equations and thermodynamics of apparent
  horizons}\/}, 2007 {\em Phys. Rev. Lett.\/} {\bf 99} 211301
  [\eprint{0704.0793}]

\bibitem{caikim}
Cai R~G and Kim S~P, {\em {First law of thermodynamics and Friedmann equations
  of Friedmann-Robertson-Walker universe}\/}, 2005 {\em JHEP\/} {\bf 02} 050
  [\eprint{hep-th/0501055}]

\bibitem{wwy}
Wu S~F, Wang B and Yang G~H, {\em {Thermodynamics on the apparent horizon in
  generalized gravity theories}\/}, 2008 {\em Nucl. Phys.\/} {\bf B799}
  330--344 [\eprint{0711.1209}]

\bibitem{duff}
Duff M~J, {\em {Twenty years of the Weyl anomaly}\/}, 1994 {\em Class. Quant.
  Grav.\/} {\bf 11} 1387--1404 [\eprint{hep-th/9308075}]

\bibitem{birrell}
Birrell N~D and Davies P~C~W, {\em {Quantum Field Theory in Curved
  Spacetime}\/}, 1982 {\em Cambridge University Press\/}

\bibitem{furs}
Fursaev D~V, {\em {Temperature and entropy of a quantum black hole and
  conformal anomaly}\/}, 1995 {\em Phys. Rev.\/} {\bf D51} 5352--5355
  [\eprint{hep-th/9412161}]

\bibitem{bm1}
Banerjee R and Majhi B~R, {\em {Quantum tunneling beyond semi-classical
  approximation}\/}, 2008 {\em JHEP\/} {\bf 06} 095 [\eprint{0805.2220}]

\bibitem{bm2}
Banerjee R and Majhi B~R, {\em {Quantum tunneling, trace anomaly and effective
  metric}\/}, 2008   [\eprint{0808.3688}]

\bibitem{cch}
Cai R~G, Cao L~M and Hu Y~P, {\em {Corrected entropy-area relation and modified
  Friedmann equations}\/}, 2008 {\em JHEP\/} {\bf 08} 090 [\eprint{0807.1232}]

\bibitem{cch2}
Cai R~G, Cao L~M and Hu Y~P, {\em {Hawking radiation of apparent horizon in a
  FRW Universe}\/}, 2008   [\eprint{0809.1554}]

\bibitem{hennsken}
Henningson M and Skenderis K, {\em {The holographic Weyl anomaly}\/}, 1998 {\em
  JHEP\/} {\bf 07} 023 [\eprint{hep-th/9806087}]

\bibitem{hartle}
Fischetti M~V, Hartle J~B and Hu B~L, {\em Quantum effects in the early
  universe {I}: Influence of trace anomalies on homogeneous, isotropic,
  classical geometries\/}, 1979 {\em Phys. Rev. D\/} {\bf 20} 1757--1771

\bibitem{hhr2}
Hawking S~W, Hertog T and Reall H~S, {\em {Trace anomaly driven inflation}\/},
  2001 {\em Phys. Rev.\/} {\bf D63} 083504 [\eprint{hep-th/0010232}]

\bibitem{kir}
Kiritsis E, {\em {Holography and brane-bulk energy exchange}\/}, 2005 {\em
  JCAP\/} {\bf 0510} 014 [\eprint{hep-th/0504219}]

\bibitem{mald}
Maldacena J~M, {\em {The large N limit of superconformal field theories and
  supergravity}\/}, 1998 {\em Adv. Theor. Math. Phys.\/} {\bf 2} 231--252
  [\eprint{hep-th/9711200}]

\bibitem{wit}
Witten E, {\em {Anti-de Sitter space and holography}\/}, 1998 {\em Adv. Theor.
  Math. Phys.\/} {\bf 2} 253--291 [\eprint{hep-th/9802150}]

\bibitem{gubpoly}
Gubser S~S, Klebanov I~R and Polyakov A~M, {\em {Gauge theory correlators from
  non-critical string theory}\/}, 1998 {\em Phys. Lett.\/} {\bf B428} 105--114
  [\eprint{hep-th/9802109}]

\bibitem{rs}
Randall L and Sundrum R, {\em {An alternative to compactification}\/}, 1999
  {\em Phys. Rev. Lett.\/} {\bf 83} 4690--4693 [\eprint{hep-th/9906064}]

\bibitem{bine}
Binetruy P, Deffayet C, Ellwanger U and Langlois D, {\em {Brane cosmological
  evolution in a bulk with cosmological constant}\/}, 2000 {\em Phys. Lett.\/}
  {\bf B477} 285--291 [\eprint{hep-th/9910219}]

\bibitem{cline}
Cline J~M, Grojean C and Servant G, {\em {Cosmological expansion in the
  presence of extra dimensions}\/}, 1999 {\em Phys. Rev. Lett.\/} {\bf 83} 4245
  [\eprint{hep-ph/9906523}]

\bibitem{maeda}
Shiromizu T, Maeda K and Sasaki M, {\em {The Einstein equations on the 3-brane
  world}\/}, 2000 {\em Phys. Rev.\/} {\bf D62} 024012 [\eprint{gr-qc/9910076}]

\bibitem{hooft}
't~Hooft G, {\em {Dimensional reduction in quantum gravity}\/}, 1993
  [\eprint{gr-qc/9310026}]

\bibitem{suss}
Susskind L, {\em {The world as a hologram}\/}, 1995 {\em J. Math. Phys.\/} {\bf
  36} 6377--6396 [\eprint{hep-th/9409089}]

\bibitem{gub}
Gubser S~S, {\em {AdS/CFT and gravity}\/}, 2001 {\em Phys. Rev.\/} {\bf D63}
  084017 [\eprint{hep-th/9912001}]

\bibitem{hhr}
Hawking S~W, Hertog T and Reall H~S, {\em {Brane new world}\/}, 2000 {\em Phys.
  Rev.\/} {\bf D62} 043501 [\eprint{hep-th/0003052}]

\bibitem{barv}
Barvinsky A~O, Deffayet C and Kamenshchik A~Y, {\em {Anomaly-driven cosmology:
  big boost scenario and AdS/CFT correspondence}\/}, 2008 {\em JCAP\/} {\bf
  0805} 020 [\eprint{0801.2063}]

\end{thebibliography}

\end{document}